\documentclass[preprint,showpacs,preprintnumbers,superscriptaddress,amsmath,amssymb,pre]{revtex4}

\usepackage{graphicx}% Include figure files
\usepackage{dcolumn}% Align table columns on decimal point
\usepackage{bm}% bold math
\usepackage{epsfig}% bold math
\newcommand \be{\begin{equation}}
\newcommand \ba{\begin{eqnarray}}
\newcommand \ea{\end{eqnarray}}
\newcommand \ee{\end{equation}}

\begin{document}

\preprint{Inverse statistics in turbulence}

\title{Multifractality of Inverse Statistics of Exit Distances in 3D
Fully Developed Turbulence}

\author{Wei-Xing Zhou}
\email{wxzhou@moho.ess.ucla.edu} \affiliation{State Key Laboratory
of Chemical Reaction Engineering,\\ East China University of
Science and Technology, Shanghai 200237, China}
\affiliation{Institute of Geophysics and Planetary Physics,
University of California, Los Angeles, CA 90095}

\author{Didier Sornette}

\email{sornette@moho.ess.ucla.edu} \affiliation{Institute of
Geophysics and Planetary Physics, University of California, Los
Angeles, CA 90095} \affiliation{Department of Earth and Space
Sciences, University of California, Los Angeles, CA 90095}
\affiliation{Laboratoire de Physique de la Mati\`ere Condens\'ee,
CNRS UMR 6622 and Universit\'e de Nice-Sophia Antipolis, 06108
Nice Cedex 2, France}

\author{Wei-Kang Yuan}
\email{wkyuan@ecust.edu.cn}\affiliation{State Key Laboratory of
Chemical Reaction Engineering,\\ East China University of Science
and Technology, Shanghai 200237, China}

\date{\today}

\begin{abstract}
The inverse structure functions of exit distances have been
introduced as a novel diagnostic of turbulence which emphasizes
the more laminar regions
\cite{Jensen-1999-PRL,Roux-Jensen-2004-PRE,Biferale-Cencini-Lanotte-Vergni-Vulpiani-2001-PRL,Biferale-Cencini-Lanotte-Vergni-2003-PF}.
Using Taylor's frozen field hypothesis, we investigate the
statistical properties of the exit distances of empirical 3D fully
developed turbulence. We find that the probability density
functions of exit distances at different velocity thresholds can
be approximated by stretched exponentials with exponents varying
with the velocity thresholds below a critical threshold. We show
that the inverse structure functions exhibit clear extended
self-similarity (ESS). The ESS exponents $\xi(p,2)$ for small $p$
($p<3.5$) are well captured by the prediction of $\xi(p,2)= p/2$
obtained by assuming a universal distribution of the exit
distances, while the observed deviations for large $p$'s
characterize the dependence of these distributions on the velocity
thresholds. By applying a box-counting multifractal analysis of
the natural measure constructed on the time series of exit
distances, we demonstrate the existence of a genuine
multifractality, endowed in addition with negative dimensions.
Performing the same analysis of reshuffled time series with
otherwise identical statistical properties for which
multifractality is absent, we show that multifractality can be
traced back to non-trivial dependence in the time series of exit
times, suggesting a non-trivial organization of weakly-turbulent
regions.
\end{abstract}


\maketitle

\section{Introduction}
\label{sec:intro}

In isotropic turbulence, structure functions are among the
favorite statistical indicators of intermittency.
The (longitudinal) structure function of order $p$ is defined by
$S_p({r})\equiv \langle \delta{v}_{\parallel}({r})^p\rangle$. The
   K41 theory \cite{KOLMOGOROV-1941-DAN} obtains that
$S_p({r}) = C_p\epsilon^{p/3}{r}^{p/3}$, where $\epsilon$ is the
average energy dissipation rate of the fluid element of size ${r}$
and $C_p$ is a constant independent of Reynolds number. The K62
theory \cite{KOLMOGOROV-1962-JFM} extends K41 by assuming a log-normal
distribution of $\epsilon$, which was questioned by Mandelbrot
\cite{Mandelbrot-1972}. The anomalous scaling properties was
uncovered experimentally
\cite{Anselmet-Gagne-Hopfinger-Antonia-1984-JFM} implying the
non-Gaussianity of the probability distribution of the velocity
increments.

The velocity structure functions consider the moments of velocity
increments over space. However, when one turns to the scalar
statistics in passive scalar advection, one often considers
averages of the advection time versus the distance
\cite{Frisch-Mazzino-Vergassola-1998-PRL,Gat-Procaccia-Zeitak-1998-PRL}.
An alternative quantity was introduced, denoted the distance
structure functions \cite{Jensen-1999-PRL} or inverse structure
functions
\cite{Biferale-Cencini-Vergni-Vulpiani-1999-PRE,Biferale-Cencini-Lanotte-Vergni-Vulpiani-2001-PRL}:
\be T_p(\delta v) \equiv \langle{{r}^p(\delta v)}\rangle,
\label{Eq:ISF1} \ee where $\delta v$ are a set of pre-chosen
thresholds of velocity increments and ${r}(\delta v)$ is the
\textit{exit distance} defined as the minimal distance for the
velocity difference to exceed $\delta{v}$
\begin{equation}
{r}(\delta v) = \inf\left\{r:|v_{i\pm r}-v_i|> \delta{v}\right\}
~, \label{Eq:rdv} \end{equation} given a record of velocity $v_i$.
In the literature, alternative definitions are adopted as well,
such as $|v_{i+r}-v_i| > \delta{v}$ or $v_{i+r}-v_i > \delta{v}$.

To ensure that the exit distance is defined, the threshold $\delta
v$ should be less than $\delta v_{\max}=(v_{\max}-v_{\min})/2$,
where $v_{\max}$ and $v_{\min}$ are respectively the maximum and
minimum of the record. On the other hand, there is a minimal
velocity increment $\delta v_{\min}=\min(|v_{i+1}-v_i|)$ for a
given record such that for any $\delta v \leq \delta v_{\min}$ we
have ${r} = 1$. Therefore, we consider the range
$(\delta{v}_{\min},\delta{v}_{\max})$. For any $\delta{v}$ in this
range, by construction, we will obtain finite ${r}$ values from the
velocity record.

The statistical properties studied for synthetic data of 24630
situations from the GOY shell model of turbulence exhibit perfect
scaling dependence of the inverse structure functions on the
velocity threshold \cite{Jensen-1999-PRL}. A completely different
result was obtained in
\cite{Biferale-Cencini-Vergni-Vulpiani-1999-PRE} where an
experimental signal was analyzed and no clear scaling was found in
the exit distance structure functions. For smoother stochastic
fluctuations associated with a spectrum with exponent $3 \leq
\beta <5$, such as two-dimensional turbulence, the inverse
structure functions exhibit bifractality
\cite{Biferale-Cencini-Lanotte-Vergni-Vulpiani-2001-PRL}. While
the large $\delta{v}$'s at fixed ${r}$ of the velocity structure
functions emphasize the most intermittent region in turbulence,
the large ${r}$'s at fixed $\delta{v}$ probe the laminar regions.
Hence, the inverse structure functions provide probes of the
intermediate dissipation range (IDR)
\cite{Biferale-Cencini-Vergni-Vulpiani-1999-PRE} introduced in
\cite{Frisch-Vergassola-1991-EPL}. It is clear that the extreme
events in the distribution of ${r}$ provide the prevailing
contributions to the inverse structure functions for large
exponents, which should thus be investigated carefully.

To our knowledge, inverse structure functions (or equivalently the statistics
of exit distances) have not been used to characterize
experimental three-dimensional turbulence data.
Here, we describe in detail the probability
distribution of exit distances ${r}$ and find that the stretched exponential
distribution is a good approximation for all ${r}$'s. Then, we analyze
the convergence of the inverse structure functions and investigate
their multiscaling properties. We
construct a measure based on the exit distance at each level
$\delta{v}$ and unveil the multifractal nature of the measure.

\section{Standard preliminary tests on the experimental data
\label{sec:expt}}

Very good quality high-Reynolds turbulence data have been
collected at the S1 ONERA wind tunnel by the Grenoble group from
LEGI \cite{Anselmet-Gagne-Hopfinger-Antonia-1984-JFM}. We use the
longitudinal velocity data obtained from this group.
The size of the velocity time series we
analyzed is $N \approx 1.73 \times 10^7$.

The mean velocity of the flow is approximately $\langle{v}\rangle
= 20 $m/s (compressive effects are thus negligible). The
root-mean-square velocity fluctuations is $v_{\mathtt{rms}} = 1.7
$m/s, leading to a turbulence intensity equal to $I =
{v_{\mathtt{rms}}} / {\langle{v}\rangle} = 0.0826$. This is
sufficiently small to allow for the use of Taylor's frozen flow
hypothesis. The integral scale is approximately $4 \mathtt{m}$ but
is difficult to estimate precisely as the turbulent flow is
neither isotropic nor homogeneous at these large scales.

The Kolmogorov microscale $\eta$ is given by
\cite{Meneveau-Sreenivasan-1991-JFM} $\eta = \left[\frac{\nu^2
\langle{v}\rangle^2}{15 \langle(\partial v/\partial t)^2\rangle
}\right]^{1/4} = 0.195 \mathtt{mm}$, where $\nu = 1.5 \times
10^{-5} \mathtt{m^2 s^{-1}}$ is the kinematic viscosity of air.
$\partial v/\partial t$ is evaluated by its discrete approximation
with a time step increment $\partial t = 3.5466 \times 10^{-5}
\mathtt{s}$ corresponding to the spatial resolution $\delta_{r} =
0.72 \mathtt{mm}$ divided by $\langle{v}\rangle$.

The Taylor scale is given by \cite{Meneveau-Sreenivasan-1991-JFM}
$\lambda =\frac{\langle{v}\rangle v_{\mathtt{rms}}}{\langle
(\partial v/\partial t)^2 \rangle^{1/2}} =16.6 \mathtt{mm}$. The
Taylor scale is thus about $85$ times the Kolmogorov scale. The
Taylor-scale Reynolds number is $Re_\lambda =
\frac{v_{\mathtt{rms}}\lambda}{\nu} = 2000$. This number is
actually not constant along the whole data set and fluctuates by
about $20\%$.

We have checked that the standard scaling laws
previously reported in the literature are recovered with this
time series. In particular, we have verified the validity
of the power-law scaling $E(k) \sim k^{-\beta}$ with an exponent
$\beta$ very close to $\frac{5}{3}$ over a range more than two
decades, similar to Fig. 5.4 of \cite{Frisch-1996} provided by
Gagne and Marchand on a similar data set from the same
experimental group. Similarly, we have checked carefully the
determination of the inertial range by combining the scaling
ranges of several velocity structure functions (see Fig. 8.6 of
\cite[Fig. 8.6]{Frisch-1996}). Conservatively, we are led to a
well-defined inertial range $60 \leq {r}/\eta \leq 2000$.

\section{Scaling properties of inverse structure functions}

\subsection{The probability distributions of exit distances}
\label{s2:PDF}

\begin{figure}
\begin{center}
\includegraphics[width=8cm, height=6cm]{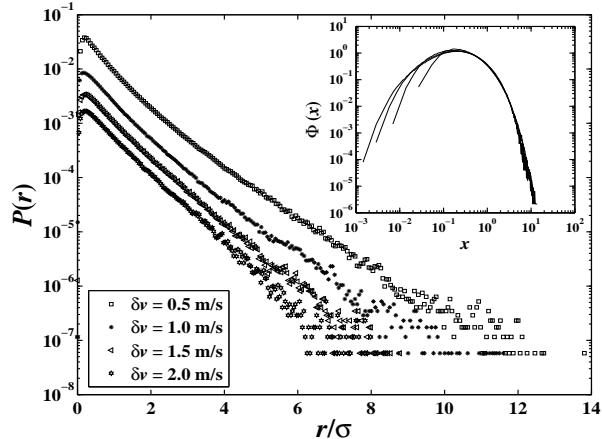}
\end{center}
\caption{Empirical probability density function $P({r})$ as a
function of normalized exit distance ${r}/\sigma$ for $\delta v =
0.5$, $1.0$, $1.5$, and $2.0$ m/s. The inset shows the
   dependence of the corresponding $\Phi(x)$ with respect to $x=r/\sigma$
   defined by (\protect\ref{Eq:Phix}).}
\label{Fig:PDF}
\end{figure}

We have obtained the exit times for 26 $\delta{v}$ values: 0.01,
0.0178, 0.0316, 0.0562, 0.1, 0.2, 0.3, 0.4, 0.5, 0.6, 0.7, 0.8,
0.9, 1, 1.1, 1.2, 1.3, 1.4, 1.5, 1.6, 1.7, 1.8, 1.9, 2, 2.33, and
2.7144 m/s. Fig.~\ref{Fig:PDF} shows the probability
density functions (pdf's) $P(r)$ as a function of $r/\sigma$ for
different velocity thresholds $\delta v = 0.5 $m/s ($\Box$),
$\delta v = 1 $m/s ($\ast$), $\delta v = 1.5 $m/s
($\triangleleft$) and $\delta v = 2 $m/s ($\star$), where
$\sigma^2=\langle r^2\rangle$ is a function of $\delta{v}$.
It is natural to normalize the exit distances by their standard
deviation $\sigma(\delta v)$ for a given $\delta v$ and
obtain the pdf of these normalized exit distances
$x=r/\sigma$ as
\begin{equation}
\Phi(x) = \sigma P(x\sigma)~. \label{Eq:Phix}
\end{equation}
The inset of Fig.~\ref{Fig:PDF} plots the corresponding
$\Phi(x)$ for the four $\delta v$ values. One can observe
an approximate collapse for $0.1 \leq x \leq 10$ but with
increasing deviations for large $x$'s. This is due to the
fact that the pdf's $P(r)$ for large $\delta{v}$ in the
semi-logarithmic plot exhibit approximate linear behaviors over a
broad range of
the normalized exit distances (exponential distribution),
while small $\delta{v}$'s have their pdf's with
fatter tail (stretched exponential distribution). Thus, the pdf's of
exit distances are not entirely described by the single scale
$\sigma(\delta v)$
but are in addition slowly varying in this structure as a function of
$\delta v$. We propose to parameterize the shape of the pdf's as
\begin{equation}
\Phi(x) = A \exp[-(x/x_0)^m]~,~~{\rm for}~x \gtrsim 1
\label{Eq:StrExpFit}
\end{equation}
where the exponent $m$ is a function of $\delta v$.
This is quite different from the inverse statistics extracted
from the time series of financial returns, for which
the distributions of exit times have power law tails
\cite{Simonsen-Jensen-Johansen-2002-EPJB,Jensen-Johansen-Simonsen-2003-PA,Jensen-Johansen-Simonsen-2003-IJMPC,Jensen-Johansen-Petroni-Simonsen-2004-PA,Zhou-Yuan-2004-XXX1}.
Actually, stretched exponential distribution is ubiquitous in
natural and social sciences, exhibiting ``fat tails'' (slower
decaying than exponential) with characteristic scales $x_0$
\cite{Laherrere-Sornette-1998-EPJB}, while power law distributions
have fat tails and are scale-free.

\begin{figure}
\begin{center}
\includegraphics[width=8cm, height=6cm]{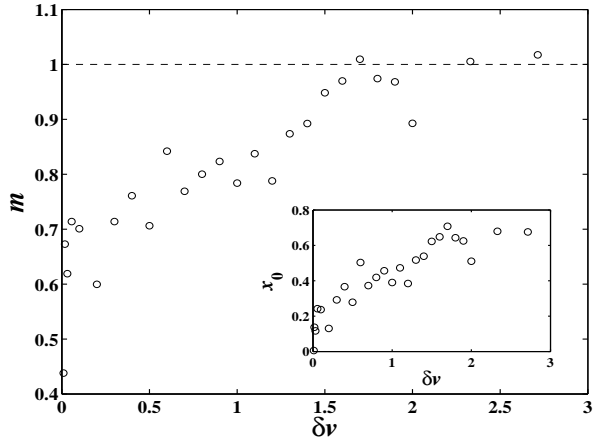}
\end{center}
\caption{The dependence of the fitted exponents $m$ and
characteristic scales $x_0$ (inset) with respect to $\delta{v}$.}
\label{Fig:m}
\end{figure}

Figure \ref{Fig:m} presents the
fitted exponents $m$ and the characteristic scales $x_0$ as a function of
$\delta{v}$. Two regimes are observed.
\begin{itemize}
\item For $\delta v \lesssim 1.5$, the exponent $m$ increases
approximately linearly with $\delta v$ as $m \approx 0.20 \times
\delta v + 0.63$. \item For $\delta v \gtrsim 1.5$, $m$ is
approximately constant with a value compatible with $m=1$
corresponding to pure exponential distributions.
\end{itemize}

\subsection{Convergence of ${x}^p\Phi(x)$} \label{s2:Converge}

\begin{figure}
\begin{center}
\includegraphics[width=8cm, height=6cm]{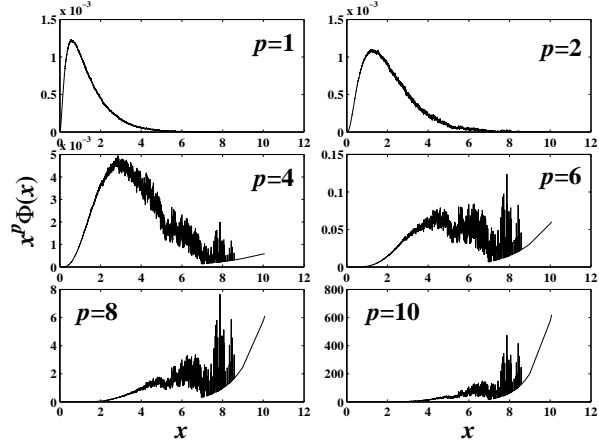}
\end{center}
\caption{Plots of ${x}^p\Phi({x})$ as a function of ${x}$ for
different values of $p$ and velocity threshold $\delta v = 1.5$
m/s. } \label{Fig:ConIntg}
\end{figure}

A preliminary condition for analyzing the inverse structure functions is
the accuracy of the moments of exit distances. One necessary
condition for $T_p$ defined by (\ref{Eq:ISF1}) to converge is that
the integrand ${r}^pP(r)$
or $x^p\Phi(x)$ converges to zero at large ${r}$, which requires
the closure of the integrand
\cite{Lvov-Podivilov-Pomyalove-Procaccia-Vandembroucq-1998-PRE}.
We have investigated $x^p\Phi(x)$ for different powers $p$ and values
$\delta{v}\in [0.01,2.7144]$ to determine how noisy is the range of $r$'s that
contribute primarily to $T_p(\delta v)$. Fixing $p$, $x^p\Phi(x)$
is more noisy for larger $\delta{v}$. For instance, $x^6\Phi(x)$
clearly converges for $\delta{v}=0.5$m/s, but not for
$\delta{v}=2.0$ m/s. We find that $T_p(\delta v)$ with $p$ up to
$5$ can be evaluated with good statistical confidence. For $p=6$,
a reasonably good evaluation of $T_6(\delta v)$ is obtained for
small $\delta{v}$. The integrands for $q \ge 8$ seem divergent and
the evaluation of the corresponding $T_p(\delta{v})$ are less
sound statistically. The typical dependence of $x^p\Phi(x)$
as a function of $x$ are shown in
Fig.~\ref{Fig:ConIntg} for $\delta v= 1.5$ m/s and $p = 1, 2, 4,
6, 8$ and $10$.

We now offer an estimate of the data size needed to estimate reliable
inverse structure functions $T_p$ for different orders $p$.
Let us assume that ${x}$ has a stretched exponential distribution
(\ref{Eq:Phix}) for ${x}$ greater than some ${x}_0$. Thus, the
integrand of $T_p$ is ${\mathcal{I}}(x) = A x^p~e^{-(x/x_0)^m}$,
where is a normalizing constant. We estimate that a reliable
estimation of $T_p$ requires a good convergence of the integrant
up to a value several times the value $x_c$ for which the integrand
achieves its maximum (we use a
factor $\kappa \approx 2-3$ according to Fig.~\ref{Fig:ConIntg}).
For the form of the stretched exponential distribution (\ref{Eq:StrExpFit}),
we have ${x}_c = x_0 (p/m)^{1/m}$. On the other hand, the largest
typical value $x_{\rm max}$ observed in a sample of size $N$ is determined
by the standard condition
\begin{equation}
N \int_{{r}_{\max}}^{\infty} \Phi(x)dx \simeq 1, \label{Eq:ellmax}
\end{equation}
where the integration can be performed analytically
in terms of a Whittaker $M$ function.
When $\Phi(x)$ is exponential, expression (\ref{Eq:ellmax})
leads to the simple equation
\begin{equation}
N = (m / A) e^{m {x}_{\rm max}}~. \label{Eq:NvsELLmax}
\end{equation}
We now write that $T_p(\delta{v})$ is reasonably well-estimated
if the range of $x$ extends at least $\kappa$ times beyond $x_c$. This
amounts to the condition
${x}_{\rm max} = \kappa {x}_c$, where $\kappa$ is approximately
independent of $\delta{v}$. It follows that the minimum sample size
necessary to calculate $T_p(\delta{v})$ for an exponential
distribution of $x$'s ($m=1$) is given by
\begin{equation}
N(p) = (m/A) e^{\kappa p}~. \label{Eq:Nvsp}
\end{equation}
For $\kappa \simeq 2$ as suggested from the left-middle panels
of Fig.~\ref{Fig:ConIntg}, with $A \approx 1/x_0$ with $x_0 \approx 0.7$
(see Fig.~\ref{Fig:m}), we find $N(p=8) \approx 6 \cdot 10^{6}$ and
$N(p=9) \approx 5 \cdot 10^{7}$. Thus, our data set with $N \approx 1.7
\times 10^7$ data points should allow us to get a reasonable estimate of
the $8$th order structure function but higher-orders become unreliable.

\subsection{Extended self-similarity of inverse structure functions}
\label{s2:ESS}

To investigate the
scaling properties of the inverse structure functions, we define a
set of relative exponents using the framework of extended self-similarity (ESS)
\cite{Benzi-Ciliberto-Tripiccione-Baudet-Massaioli-Succi-1993-PRE}:
\be
T_p(\delta v) \propto [T_{p_0}(\delta v)]^{\xi(p,p_0)},
\label{Eq:Defxi01}
\ee
where $p_0$ is a reference order. In the
case of velocity structure functions, $p_0=3$ is a quite natural
choice based on the exact Kolmogorov's four-fifth law.
There is no similar reference for the scaling properties of
the inverse structure functions and we choose somewhat
arbitrarily $p_0=2$. In general, ESS provides a
wider scaling range for the extraction of scaling exponents. We
will see in this subsection that Eq.~(\ref{Eq:Defxi01}) holds
for our experimental data of turbulence with a high accuracy.

\begin{figure}[htb]
\begin{center}
\epsfig{file=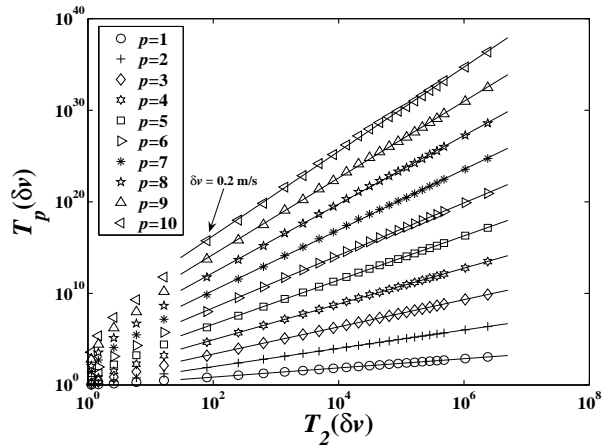,width=8cm, height=6cm}
\end{center}
\caption{Dependence of the inverse structure function of order $p$
as a function of the inverse structure function of order
$2$ taken as a reference. The straight lines exemplify the extended
self-similarity of
inverse structure functions for the threshold $\delta v$ ranging
from $0.2 $m/s to $2.7144 $m/s.} \label{Fig:TpvsT2}
\end{figure}

Figure~\ref{Fig:TpvsT2} presents log-log plots of $T_p(\delta v)$
vs $T_2(\delta v)$ for $p = 1, 2, \cdots, 10$ with $\delta{v} \in
[0.01, 2.7144]$. The straight lines hold for $0.2 \leq \delta v
\leq 2.7144$ and over at least four orders of magnitudes in
$T_2(\delta v)$, showing the existence of extended self-similarity
in the inverse structure functions. The scaling range for small
$p$'s seems to be broader than for large $p$'s.

\begin{figure}[htb]
\begin{center}
\includegraphics[width=8cm, height=6cm]{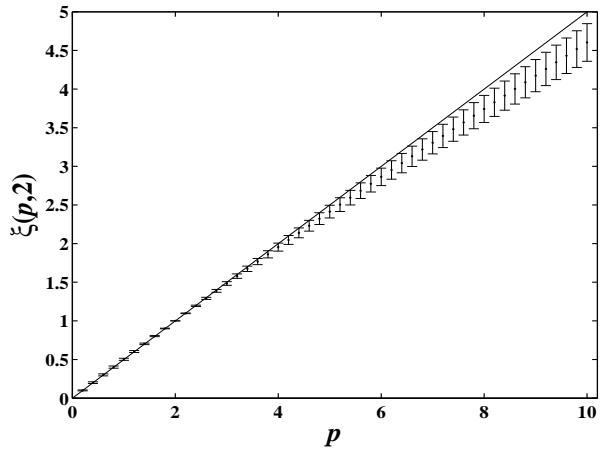}
\end{center}
\caption{Dependence of the ESS exponents $\xi(p,2)$ as a function
of the order $p$ of the inverse structure functions. The straight
line is the prediction (\ref{Eq:xip2}) obtained under the
assumption of a universal pdf $\Phi_{\delta v}(x)$ of the exit
distances (which is independent of the velocity thresholds $\delta
v$).} \label{Fig:XI}
\end{figure}

The ESS scaling exponents $\xi(p,2)$ are shown in
Fig.~\ref{Fig:XI}. The error bars on  $\xi(p,2)$ corresponds to
$\pm$ one standard deviation. There is an indication that
$\xi(p,2)$ has a nonlinear dependence as a function of $p$, with a
downward curvature making the curve depart from the linear
dependence $\xi(p,2) = p/2$ observes for small $p$'s.

The monoscaling behavior
\begin{equation}
   \xi(p, 2) = p/2 \label{Eq:xip2}
\end{equation}
is predicted from the assumption that the pdf $\Phi(x)$ of the
normalized exit distances, and given by (\ref{Eq:StrExpFit}), is
independent of $\delta v$. By definition, we have $T_p(\delta{v})
= \langle{{r}^p}\rangle_{\delta{v}} = \int_{0}^{\infty} {r}^p
P_{\delta v}({r})d{r} = [\sigma(\delta v)]^p \int_{0}^{\infty}
dx~\Phi_{\delta v}(x) x^p$. Thus, \be T_p(\delta{v}) =
[T_2(\delta{v})]^{p/2}~ \frac{\int_{0}^{\infty} dx~\Phi_{\delta
v}(x) ~x^p}{\left[\int_{0}^{\infty} dx~\Phi_{\delta v}(x) ~x^2
\right]^{p/2}}~. \label{mngjlsls} \ee If $\Phi_{\delta v}(x)$ is
universal and independent of $\delta v$, then the last term in
(\ref{mngjlsls}) is a number independent of $\delta v$ (and thus
of $T_2(\delta{v})$) and the mono-scaling $\xi(p,2) = p/2$ follows.
Thus, the prediction  (\ref{Eq:xip2}) holds for those velocity
thresholds $\delta{v}$ satisfying the condition that the pdf of
exit distance is universal (independent of $\delta v$). This is
the analog of the K41 prediction on the standard structure
functions. In our present case, there are deviations of the pdf's
of exit distances from the exponential law at small $\delta{v}$,
that we have proposed to be quantified under the form of stretched
exponentials (\ref{Eq:StrExpFit}) with exponents $m(\delta v)$
being a function of $\delta v$ as shown in Fig.~\ref{Fig:m}. These
deviations from exact self-similarity are weaker than for the
direct statistics and are revealed more clearly for the higher
orders. We can therefore attribute the deviation of the empirical
$\xi(p,2)$ from the self-similarity (\ref{Eq:xip2}) at high orders
to the non-universality of $\Phi_{\delta v}(x)$ which depends on
the velocity levels $\delta{v}$.

\section{Multifractality of the time series of exit distances at different
$\delta{v}$ \label{s1:MF}}

To investigate further the multifractal nature of the exit
distance series $\{r(t):t=1,\cdots,N = 3\times 11 \times 2^{19}
\approx 1.7 \times 10^7\}$ for a given $\delta{v}$, we construct a
probability measure $\mu$ through its integral function $M(t)$
\begin{equation}
\mu\left(]t_1,t_2]\right) = M(t_2) - M(t_1)~,
\label{ngjlkls,lwe}
\end{equation}
where $M(t)=0$ for $0<t<1$ and \be
M(t) = \sum_{i=1}^{[t]} r(i)
\ee
for $1\le t\le N$.

The box-counting method allows us to test for a
possible multifractality of the measure $\mu$. The sizes $s$ of the
covering boxes are chosen such that the number of boxes of
each size is an integer: $n = N/s \in\cal{N}$. We construct
the partition function $Z_q$ as
\begin{equation}
Z_q(s) \triangleq \sum_{i=1}^n
\left[\mu\left(](i-1)s,is]\right)\right]^q~. \label{Eq:Zq}
\end{equation}
and expect it to scale as
\cite{Frisch-Parisi-1985,Halsey-Jensen-Kadanoff-Procaccia-Shraiman-1986-PRA}
\be Z_q(s) \sim s^{\tau(q)}~, \label{Eq:Tau} \ee which defines the
exponent $\tau(q)$. For $\tau(q)$, a hierarchy of generalized
dimensions $D_q$
\cite{Grassberger-1983-PLA,Hentschel-Procaccia-1983-PD,Grassberger-Procaccia-1983-PD}
can be calculated according to
\begin{equation}
D_q = \lim_{p\to q} \frac{\tau(p)}{p-1}~. \label{Eq:Dq}
\end{equation}
$D_0$ is the fractal dimension of the support of the measure. For
our measure (\ref{ngjlkls,lwe}), we have $D_0=1$.
The local singularity exponent $\alpha$ of the measure $\mu$ and its spectrum
$f(\alpha)$ are related to $\tau(q)$ through a Legendre
transformation
\cite{Halsey-Jensen-Kadanoff-Procaccia-Shraiman-1986-PRA}
\begin{subequations}
\begin{equation}
\alpha(q) = d\tau(q)/dq~, \label{Eq:Alpha}
\end{equation}
and
\begin{equation}
f[\alpha(q)] = q\alpha(q)-\tau(q)~. \label{Eq:f}
\end{equation}
\end{subequations}

\begin{figure}[htb]
\begin{center}
\includegraphics[width=8cm,height=6cm]{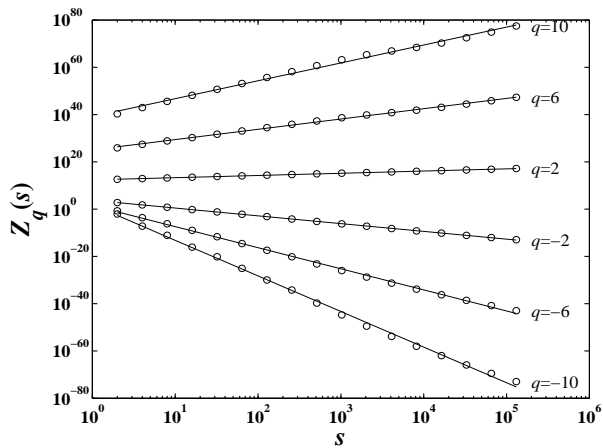}
\end{center}
\caption{Plots of $Z_q(s)$ for $\delta{v}=1.5$m/s as a function of
the box size $s$ for different values of $q$ in log-log
coordinates. The solid lines are the least-square fits to the data
using a linear regression (in log-log coordinates) corresponding
to power laws.}
\label{Fig:Zq}
\end{figure}

We have tested the power-law scaling of $Z_q(s)$ as a function of
the box size $s$ for the exit time sequences at different velocity
levels $\delta{v}$. The scaling range is found to span over four
orders of magnitude. Figure \ref{Fig:Zq} plots the partition
function $Z_q(s)$ for $\delta{v}=1.5$m/s as a function of the box
size $s$ for six different values of $q$ in log-log coordinates.
The solid lines are the least-square fits with power laws for each
$q$. The correlation coefficients of the linear regressions (in
log-log) are all larger than $0.997$, demonstrating the existence
of a very good scaling.

\begin{figure}[htb]
\begin{center}
\includegraphics[width=8cm, height=6cm]{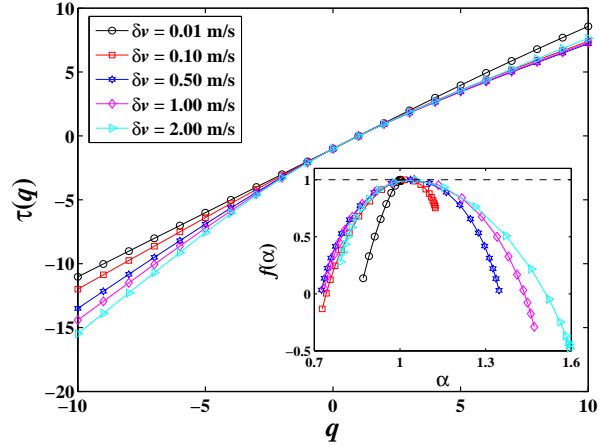}
\end{center}
\caption{(Color online) Scaling exponents $\tau(q)$ as a function
of $q$ for different velocity levels $\delta v$. The inset shows
the fractal spectra $f(\alpha)$ obtained by the Legendre transform
of $\tau(q)$.} \label{Fig:Taufalpha}
\end{figure}

The scaling exponents $\tau(q)$ are given by the slopes of the
linear fits of $\ln[Z_q(s)]$ as a function of $\ln s$ for
different values of $\delta{v}$. Figure \ref{Fig:Taufalpha} plots
$\tau(q)$ as a function of $q$ for five different velocity levels
$\delta v$. The inset shows the fractal spectra $f(\alpha)$
obtained by the Legendre transformation of $\tau(q)$ defined by
(\ref{Eq:f}). We observe that $\tau(q)$'s are concave and
nonlinear, a diagnostic of multifractality. The maximal and
minimal strength of the set of singularities, $\alpha_{\max}$ and
$\alpha_{\min}$, can be approximated asymptotically by
$\lim_{q\to\infty}D_q = \lim_{q\to\infty}\tau(q)/q$ and
$\lim_{q\to-\infty}D_q = \lim_{q\to-\infty}\tau(q)/q$,
respectively. It can be clearly observed that the steepness of the
curve $\tau(q)$ for negative $q$ increases with $\delta{v}$.
Consequently, the maximal singularity $\alpha_{\max}$ increases
with $\delta{v}$, as shown in the inset of
Fig.~\ref{Fig:Taufalpha} where the value of $\alpha$ at the right
endpoint increases with $\delta{v}$.

Since $\mu$ is conservative, $\tau(1) = 0$ and $\tau(0)=-1$ for
all $\delta{v}$. For a given $q<-1$, the function $\tau(q)$
decreases with $\delta{v}$. For a given $q>1$, there is a critical
value $\delta{v_c}$ such that $d\tau(q)/d\delta{v}<0$ when
$\delta{v}<\delta{v_c}$ and $d\tau(q)/d\delta{v}>0$ when
$\delta{v}>\delta{v_c}$. We find that $\delta{v}_c$ can be
approximated by a linear function
\begin{equation}
\delta{v}_c = -0.058 q + 0.783~, \label{Eq:dvc}
\end{equation}
associated with a correlation coefficient of the linear regression equal to
$0.963$. In addition, one can also see that $\alpha_{\min}$
decreases with $\delta v$ for $\delta{v}<\delta{v_c}$ and increases with
$\delta v$ for
$\delta{v}>\delta{v_c}$. For $q=-1$, we find for instance that $\delta{v_c}
\approx 1.2 \rm{m}/\rm{s}$.

We have performed exactly the same multifractal analysis as done
above on synthetic time series generated from a stretched
exponential distribution and on reshuffled data of the real exit
distances. Both tests give linear scaling exponents $\tau(q)=q-1$
in a narrower scaling range $64 \le s \le 131072$, which is the
earmark of monofractality. These tests strengthen  the presence of
multifractality extracted from the real exit distance data.

In general, multifractality in time series can be attributed to
either a hierarchy of changing broad probability density functions
for the values of the time series at different scales and/or
different long-range temporal correlations of the small and large
fluctuations
\cite{Kantelhardt-Zschiegner-Bunde-Havlin-Bunde-Stanley-2002-PA}.
Our comparison, with reshuffled data and sequences with the same
pdf's but no correlation which exhibit trivial monofractality,
suggests that multifractality in the set of exit distances may be
attributed at least in part to the existence non-trivial
dependence in the time series of exit distances.

An important feature of the multifractal spectrum $f(\alpha)$ in
the inset of Fig.~\ref{Fig:Taufalpha} is the existence of negative
(or latent) dimensions, that is, $f(\alpha)<0$
\cite{Mandelbrot-1989,Mandelbrot-1990-PA,Mandelbrot-1991-PRSLA,Chhabra-Sreenivasan-1991-PRA}.
The source of negative dimensions could be twofold. Firstly, the
turbulent flow is a stochastic process, which introduces intrinsic
randomness in the multifractal measure $\mu$. We note that
negative dimensions also appear in continuous multifractals
\cite{Zhou-Liu-Yu-2001-Fractals,Zhou-Yu-2001-PA,Zhou-Yu-2001-PRE}.
Secondly, negative dimensions may be interpreted geometrically by
considering cuts of higher dimensional multifractals
\cite{Mandelbrot-1989,Mandelbrot-1990-PA,Mandelbrot-1991-PRSLA}.
This intuition proposed by Mandelbrot has been proved mathematically in the
multifractal slice theorem
\cite{Olsen-1998-PMH,Olsen-1999-HMJ,Olsen-2000-PP}. In the present
case, the frozen field hypothesis is applied and we deal with
one-dimensional cut of the three dimensional turbulence velocity
field.

\section{Concluding remarks}

Based on Taylor's frozen field hypothesis, the statistical
properties of the exit distances of 3D turbulence have been
investigated. The probability density functions of exit distances
at different velocity thresholds have been shown to be well
approximated by stretched exponentials. The inverse structure
functions was shown to exhibit very clear extended self-similarity
(ESS). The ESS exponents $\xi(p,2)$ for small $p<3.5$ are well
described by the monofractal prediction $\xi(p,2)= p/2$ obtained
by assuming a universal exponential distribution of the exit
distance. The multifractality is thus related to the dependence of
the pdf's of the normalized exit distances on the velocity
thresholds $\delta v$. We have demonstrated that the sequences of
exit distances for each velocity threshold $\delta{v}$ exhibit a
clear multifractality with negative dimensions. The scaling ranges
over which multifractality holds cover more than four order of
magnitude in the exit distance variable. The comparison, with
reshuffled data and sequences with the same pdf's but no
correlation which exhibit trivial monofractality, suggests
strongly that our report of multifractality is not artifactual.

Our report of multifractality in the time series of exit distance,
which tends to emphasize the least turbulent/most laminar regions,
suggests a much richer organization of the weakly turbulent and close
to laminar regions than believed until recently.

\begin{acknowledgements}
The research by Zhou and Yuan was supported by NSFC/PetroChina
jointly through a major project on multiscale methodology (No.
20490200).
\end{acknowledgements}

\bibliography{InverseSF}

\begin{thebibliography}{38}
\expandafter\ifx\csname natexlab\endcsname\relax\def\natexlab#1{#1}\fi
\expandafter\ifx\csname bibnamefont\endcsname\relax
  \def\bibnamefont#1{#1}\fi
\expandafter\ifx\csname bibfnamefont\endcsname\relax
  \def\bibfnamefont#1{#1}\fi
\expandafter\ifx\csname citenamefont\endcsname\relax
  \def\citenamefont#1{#1}\fi
\expandafter\ifx\csname url\endcsname\relax
  \def\url#1{\texttt{#1}}\fi
\expandafter\ifx\csname urlprefix\endcsname\relax\def\urlprefix{URL }\fi
\providecommand{\bibinfo}[2]{#2}
\providecommand{\eprint}[2][]{\url{#2}}

\bibitem[{\citenamefont{Jensen}(1999)}]{Jensen-1999-PRL}
\bibinfo{author}{\bibfnamefont{M.~H.} \bibnamefont{Jensen}},
  \bibinfo{journal}{Phys. Rev. Lett.} \textbf{\bibinfo{volume}{83}},
  \bibinfo{pages}{76} (\bibinfo{year}{1999}).

\bibitem[{\citenamefont{Roux and Jensen}(2004)}]{Roux-Jensen-2004-PRE}
\bibinfo{author}{\bibfnamefont{S.}~\bibnamefont{Roux}} \bibnamefont{and}
  \bibinfo{author}{\bibfnamefont{M.~H.} \bibnamefont{Jensen}},
  \bibinfo{journal}{Phys. Rev. E} \textbf{\bibinfo{volume}{69}},
  \bibinfo{pages}{016309} (\bibinfo{year}{2004}).

\bibitem[{\citenamefont{Biferale et~al.}(2001)\citenamefont{Biferale, Cencini,
  Lanotte, Vergni, and
  Vulpiani}}]{Biferale-Cencini-Lanotte-Vergni-Vulpiani-2001-PRL}
\bibinfo{author}{\bibfnamefont{L.}~\bibnamefont{Biferale}},
  \bibinfo{author}{\bibfnamefont{M.}~\bibnamefont{Cencini}},
  \bibinfo{author}{\bibfnamefont{A.~S.} \bibnamefont{Lanotte}},
  \bibinfo{author}{\bibfnamefont{D.}~\bibnamefont{Vergni}}, \bibnamefont{and}
  \bibinfo{author}{\bibfnamefont{A.}~\bibnamefont{Vulpiani}},
  \bibinfo{journal}{Phys. Rev. Lett.} \textbf{\bibinfo{volume}{87}},
  \bibinfo{pages}{124501} (\bibinfo{year}{2001}).

\bibitem[{\citenamefont{Biferale et~al.}(2003)\citenamefont{Biferale, Cencini,
  Lanotte, and Vergni}}]{Biferale-Cencini-Lanotte-Vergni-2003-PF}
\bibinfo{author}{\bibfnamefont{L.}~\bibnamefont{Biferale}},
  \bibinfo{author}{\bibfnamefont{M.}~\bibnamefont{Cencini}},
  \bibinfo{author}{\bibfnamefont{A.~S.} \bibnamefont{Lanotte}},
  \bibnamefont{and} \bibinfo{author}{\bibfnamefont{D.}~\bibnamefont{Vergni}},
  \bibinfo{journal}{Phys. Fluids} \textbf{\bibinfo{volume}{15}},
  \bibinfo{pages}{1012} (\bibinfo{year}{2003}).

\bibitem[{\citenamefont{Kolmogorov}(1941)}]{KOLMOGOROV-1941-DAN}
\bibinfo{author}{\bibfnamefont{A.~N.} \bibnamefont{Kolmogorov}},
  \bibinfo{journal}{Dokl. Akad. Nauk SSSR} \textbf{\bibinfo{volume}{30}},
  \bibinfo{pages}{9} (\bibinfo{year}{1941}), \bibinfo{note}{(reprinted in Proc.
  R. Soc.Lond. A 434, 15-17 (1991))}.

\bibitem[{\citenamefont{Kolmogorov}(1962)}]{KOLMOGOROV-1962-JFM}
\bibinfo{author}{\bibfnamefont{A.~N.} \bibnamefont{Kolmogorov}},
  \bibinfo{journal}{J. Fluid Mech.} \textbf{\bibinfo{volume}{13}},
  \bibinfo{pages}{82} (\bibinfo{year}{1962}).

\bibitem[{\citenamefont{Mandelbrot}(1972)}]{Mandelbrot-1972}
\bibinfo{author}{\bibfnamefont{B.~B.} \bibnamefont{Mandelbrot}}, in
  \emph{\bibinfo{booktitle}{Lecture Notes in Physics}}, edited by
  \bibinfo{editor}{\bibfnamefont{M.}~\bibnamefont{Rosenblatt}}
  \bibnamefont{and} \bibinfo{editor}{\bibfnamefont{C.}~\bibnamefont{van Atta}}
  (\bibinfo{publisher}{Springer}, \bibinfo{year}{1972}),
  vol.~\bibinfo{volume}{12}, pp. \bibinfo{pages}{333--351}.

\bibitem[{\citenamefont{Anselmet et~al.}(1984)\citenamefont{Anselmet, Gagne,
  Hopfinger, and Antonia}}]{Anselmet-Gagne-Hopfinger-Antonia-1984-JFM}
\bibinfo{author}{\bibfnamefont{F.}~\bibnamefont{Anselmet}},
  \bibinfo{author}{\bibfnamefont{Y.}~\bibnamefont{Gagne}},
  \bibinfo{author}{\bibfnamefont{E.~J.} \bibnamefont{Hopfinger}},
  \bibnamefont{and} \bibinfo{author}{\bibfnamefont{R.~A.}
  \bibnamefont{Antonia}}, \bibinfo{journal}{J. Fluid Mech.}
  \textbf{\bibinfo{volume}{140}}, \bibinfo{pages}{63} (\bibinfo{year}{1984}).

\bibitem[{\citenamefont{Frisch et~al.}(1998)\citenamefont{Frisch, Mazzino, and
  Vergassola}}]{Frisch-Mazzino-Vergassola-1998-PRL}
\bibinfo{author}{\bibfnamefont{U.}~\bibnamefont{Frisch}},
  \bibinfo{author}{\bibfnamefont{A.}~\bibnamefont{Mazzino}}, \bibnamefont{and}
  \bibinfo{author}{\bibfnamefont{M.}~\bibnamefont{Vergassola}},
  \bibinfo{journal}{Phys. Rev. Lett.} \textbf{\bibinfo{volume}{80}},
  \bibinfo{pages}{5532} (\bibinfo{year}{1998}).

\bibitem[{\citenamefont{Gat et~al.}(1998)\citenamefont{Gat, Procaccia, and
  Zeitak}}]{Gat-Procaccia-Zeitak-1998-PRL}
\bibinfo{author}{\bibfnamefont{O.}~\bibnamefont{Gat}},
  \bibinfo{author}{\bibfnamefont{I.}~\bibnamefont{Procaccia}},
  \bibnamefont{and} \bibinfo{author}{\bibfnamefont{R.}~\bibnamefont{Zeitak}},
  \bibinfo{journal}{Phys. Rev. Lett.} \textbf{\bibinfo{volume}{80}},
  \bibinfo{pages}{5536} (\bibinfo{year}{1998}).

\bibitem[{\citenamefont{Biferale et~al.}(1999)\citenamefont{Biferale, Cencini,
  Vergni, and Vulpiani}}]{Biferale-Cencini-Vergni-Vulpiani-1999-PRE}
\bibinfo{author}{\bibfnamefont{L.}~\bibnamefont{Biferale}},
  \bibinfo{author}{\bibfnamefont{M.}~\bibnamefont{Cencini}},
  \bibinfo{author}{\bibfnamefont{D.}~\bibnamefont{Vergni}}, \bibnamefont{and}
  \bibinfo{author}{\bibfnamefont{A.}~\bibnamefont{Vulpiani}},
  \bibinfo{journal}{Phys. Rev. E} \textbf{\bibinfo{volume}{60}},
  \bibinfo{pages}{R6295} (\bibinfo{year}{1999}).

\bibitem[{\citenamefont{Frisch and
  Vergassola}(1991)}]{Frisch-Vergassola-1991-EPL}
\bibinfo{author}{\bibfnamefont{U.}~\bibnamefont{Frisch}} \bibnamefont{and}
  \bibinfo{author}{\bibfnamefont{M.}~\bibnamefont{Vergassola}},
  \bibinfo{journal}{Europhys. Lett.} \textbf{\bibinfo{volume}{14}},
  \bibinfo{pages}{439} (\bibinfo{year}{1991}).

\bibitem[{\citenamefont{Meneveau and
  Sreenivasan}(1991)}]{Meneveau-Sreenivasan-1991-JFM}
\bibinfo{author}{\bibfnamefont{C.}~\bibnamefont{Meneveau}} \bibnamefont{and}
  \bibinfo{author}{\bibfnamefont{K.~R.} \bibnamefont{Sreenivasan}},
  \bibinfo{journal}{J. Fluid Mech.} \textbf{\bibinfo{volume}{224}},
  \bibinfo{pages}{429} (\bibinfo{year}{1991}).

\bibitem[{\citenamefont{Frisch}(1996)}]{Frisch-1996}
\bibinfo{author}{\bibfnamefont{U.}~\bibnamefont{Frisch}},
  \emph{\bibinfo{title}{Turbulence: The Legacy of A.N. Kolmogorov}}
  (\bibinfo{publisher}{Cambridge University Press},
  \bibinfo{address}{Cambridge}, \bibinfo{year}{1996}).

\bibitem[{\citenamefont{Simonsen et~al.}(2002)\citenamefont{Simonsen, Jensen,
  and Johansen}}]{Simonsen-Jensen-Johansen-2002-EPJB}
\bibinfo{author}{\bibfnamefont{I.}~\bibnamefont{Simonsen}},
  \bibinfo{author}{\bibfnamefont{M.~H.} \bibnamefont{Jensen}},
  \bibnamefont{and} \bibinfo{author}{\bibfnamefont{A.}~\bibnamefont{Johansen}},
  \bibinfo{journal}{Eur. Phys. J. B} \textbf{\bibinfo{volume}{27}},
  \bibinfo{pages}{583} (\bibinfo{year}{2002}).

\bibitem[{\citenamefont{Jensen et~al.}(2003{\natexlab{a}})\citenamefont{Jensen,
  Johansen, and Simonsen}}]{Jensen-Johansen-Simonsen-2003-PA}
\bibinfo{author}{\bibfnamefont{M.~H.} \bibnamefont{Jensen}},
  \bibinfo{author}{\bibfnamefont{A.}~\bibnamefont{Johansen}}, \bibnamefont{and}
  \bibinfo{author}{\bibfnamefont{I.}~\bibnamefont{Simonsen}},
  \bibinfo{journal}{Physica A} \textbf{\bibinfo{volume}{324}},
  \bibinfo{pages}{338} (\bibinfo{year}{2003}{\natexlab{a}}).

\bibitem[{\citenamefont{Jensen et~al.}(2003{\natexlab{b}})\citenamefont{Jensen,
  Johansen, and Simonsen}}]{Jensen-Johansen-Simonsen-2003-IJMPC}
\bibinfo{author}{\bibfnamefont{M.~H.} \bibnamefont{Jensen}},
  \bibinfo{author}{\bibfnamefont{A.}~\bibnamefont{Johansen}}, \bibnamefont{and}
  \bibinfo{author}{\bibfnamefont{I.}~\bibnamefont{Simonsen}},
  \bibinfo{journal}{Int. J. Mod. Phys. B} \textbf{\bibinfo{volume}{17}},
  \bibinfo{pages}{4003} (\bibinfo{year}{2003}{\natexlab{b}}).

\bibitem[{\citenamefont{Jensen et~al.}(2004)\citenamefont{Jensen, Johansen,
  Petroni, and Simonsen}}]{Jensen-Johansen-Petroni-Simonsen-2004-PA}
\bibinfo{author}{\bibfnamefont{M.~H.} \bibnamefont{Jensen}},
  \bibinfo{author}{\bibfnamefont{A.}~\bibnamefont{Johansen}},
  \bibinfo{author}{\bibfnamefont{F.}~\bibnamefont{Petroni}}, \bibnamefont{and}
  \bibinfo{author}{\bibfnamefont{I.}~\bibnamefont{Simonsen}},
  \bibinfo{journal}{Physica A} \textbf{\bibinfo{volume}{340}},
  \bibinfo{pages}{678} (\bibinfo{year}{2004}).

\bibitem[{\citenamefont{Zhou and Yuan}(2004)}]{Zhou-Yuan-2004-XXX1}
\bibinfo{author}{\bibfnamefont{W.-X.} \bibnamefont{Zhou}} \bibnamefont{and}
  \bibinfo{author}{\bibfnamefont{W.-K.} \bibnamefont{Yuan}}
  (\bibinfo{year}{2004}), \bibinfo{note}{preprint at cond-mat/0410225}.

\bibitem[{\citenamefont{Laherrere and
  Sornette}(1998)}]{Laherrere-Sornette-1998-EPJB}
\bibinfo{author}{\bibfnamefont{J.}~\bibnamefont{Laherrere}} \bibnamefont{and}
  \bibinfo{author}{\bibfnamefont{D.}~\bibnamefont{Sornette}},
  \bibinfo{journal}{Eur. Phys. J. B} \textbf{\bibinfo{volume}{2}},
  \bibinfo{pages}{525} (\bibinfo{year}{1998}).

\bibitem[{\citenamefont{L'vov et~al.}(1998)\citenamefont{L'vov, Podivilov,
  Pomyalov, Procaccia, and
  Vandembroucq}}]{Lvov-Podivilov-Pomyalove-Procaccia-Vandembroucq-1998-PRE}
\bibinfo{author}{\bibfnamefont{V.~S.} \bibnamefont{L'vov}},
  \bibinfo{author}{\bibfnamefont{E.}~\bibnamefont{Podivilov}},
  \bibinfo{author}{\bibfnamefont{A.}~\bibnamefont{Pomyalov}},
  \bibinfo{author}{\bibfnamefont{I.}~\bibnamefont{Procaccia}},
  \bibnamefont{and}
  \bibinfo{author}{\bibfnamefont{D.}~\bibnamefont{Vandembroucq}},
  \bibinfo{journal}{Phys. Rev. E} \textbf{\bibinfo{volume}{58}},
  \bibinfo{pages}{1811} (\bibinfo{year}{1998}).

\bibitem[{\citenamefont{Benzi et~al.}(1993)\citenamefont{Benzi, Ciliberto,
  Tripiccione, Baudet, Massaioli, and
  Succi}}]{Benzi-Ciliberto-Tripiccione-Baudet-Massaioli-Succi-1993-PRE}
\bibinfo{author}{\bibfnamefont{R.}~\bibnamefont{Benzi}},
  \bibinfo{author}{\bibfnamefont{S.}~\bibnamefont{Ciliberto}},
  \bibinfo{author}{\bibfnamefont{R.}~\bibnamefont{Tripiccione}},
  \bibinfo{author}{\bibfnamefont{C.}~\bibnamefont{Baudet}},
  \bibinfo{author}{\bibfnamefont{F.}~\bibnamefont{Massaioli}},
  \bibnamefont{and} \bibinfo{author}{\bibfnamefont{S.}~\bibnamefont{Succi}},
  \bibinfo{journal}{Phys. Rev. E} \textbf{\bibinfo{volume}{48}},
  \bibinfo{pages}{R29} (\bibinfo{year}{1993}).

\bibitem[{\citenamefont{Frisch and Parisi}(1985)}]{Frisch-Parisi-1985}
\bibinfo{author}{\bibfnamefont{U.}~\bibnamefont{Frisch}} \bibnamefont{and}
  \bibinfo{author}{\bibfnamefont{G.}~\bibnamefont{Parisi}}, in
  \emph{\bibinfo{booktitle}{Turbulence and Predictability in Geophysical Fluid
  Dynamics}}, edited by \bibinfo{editor}{\bibfnamefont{P.~G.}
  \bibnamefont{Gil~M}, \bibfnamefont{Benzi~R}}
  (\bibinfo{publisher}{North-Holland}, \bibinfo{year}{1985}), pp.
  \bibinfo{pages}{84--88}.

\bibitem[{\citenamefont{Halsey et~al.}(1986)\citenamefont{Halsey, Jensen,
  Kadanoff, Procaccia, and
  Shraiman}}]{Halsey-Jensen-Kadanoff-Procaccia-Shraiman-1986-PRA}
\bibinfo{author}{\bibfnamefont{T.~C.} \bibnamefont{Halsey}},
  \bibinfo{author}{\bibfnamefont{M.~H.} \bibnamefont{Jensen}},
  \bibinfo{author}{\bibfnamefont{L.~P.} \bibnamefont{Kadanoff}},
  \bibinfo{author}{\bibfnamefont{I.}~\bibnamefont{Procaccia}},
  \bibnamefont{and} \bibinfo{author}{\bibfnamefont{B.~I.}
  \bibnamefont{Shraiman}}, \bibinfo{journal}{Phys. Rev. A}
  \textbf{\bibinfo{volume}{33}}, \bibinfo{pages}{1141} (\bibinfo{year}{1986}).

\bibitem[{\citenamefont{Grassberger}(1983)}]{Grassberger-1983-PLA}
\bibinfo{author}{\bibfnamefont{P.}~\bibnamefont{Grassberger}},
  \bibinfo{journal}{Phys. Lett. A} \textbf{\bibinfo{volume}{97}},
  \bibinfo{pages}{227} (\bibinfo{year}{1983}).

\bibitem[{\citenamefont{Hentschel and
  Procaccia}(1983)}]{Hentschel-Procaccia-1983-PD}
\bibinfo{author}{\bibfnamefont{H.~G.~E.} \bibnamefont{Hentschel}}
  \bibnamefont{and}
  \bibinfo{author}{\bibfnamefont{I.}~\bibnamefont{Procaccia}},
  \bibinfo{journal}{Physica D} \textbf{\bibinfo{volume}{8}},
  \bibinfo{pages}{435} (\bibinfo{year}{1983}).

\bibitem[{\citenamefont{Grassberger and
  Procaccia}(1983)}]{Grassberger-Procaccia-1983-PD}
\bibinfo{author}{\bibfnamefont{P.}~\bibnamefont{Grassberger}} \bibnamefont{and}
  \bibinfo{author}{\bibfnamefont{I.}~\bibnamefont{Procaccia}},
  \bibinfo{journal}{Physica D} \textbf{\bibinfo{volume}{9}},
  \bibinfo{pages}{189} (\bibinfo{year}{1983}).

\bibitem[{\citenamefont{Kantelhardt et~al.}(2002)\citenamefont{Kantelhardt,
  Zschiegner, Koscielny-Bunde, Havlin, Bunde, and
  Stanley}}]{Kantelhardt-Zschiegner-Bunde-Havlin-Bunde-Stanley-2002-PA}
\bibinfo{author}{\bibfnamefont{J.~W.} \bibnamefont{Kantelhardt}},
  \bibinfo{author}{\bibfnamefont{S.~A.} \bibnamefont{Zschiegner}},
  \bibinfo{author}{\bibfnamefont{E.}~\bibnamefont{Koscielny-Bunde}},
  \bibinfo{author}{\bibfnamefont{S.}~\bibnamefont{Havlin}},
  \bibinfo{author}{\bibfnamefont{A.}~\bibnamefont{Bunde}}, \bibnamefont{and}
  \bibinfo{author}{\bibfnamefont{H.~E.} \bibnamefont{Stanley}},
  \bibinfo{journal}{Physica A} \textbf{\bibinfo{volume}{316}},
  \bibinfo{pages}{87} (\bibinfo{year}{2002}).

\bibitem[{\citenamefont{Mandelbrot}(1989)}]{Mandelbrot-1989}
\bibinfo{author}{\bibfnamefont{B.~B.} \bibnamefont{Mandelbrot}}, in
  \emph{\bibinfo{booktitle}{Fractals' Physical Origin and Properties}}, edited
  by \bibinfo{editor}{\bibfnamefont{L.}~\bibnamefont{Pietronero}}
  (\bibinfo{publisher}{Plenum}, \bibinfo{address}{New York},
  \bibinfo{year}{1989}), pp. \bibinfo{pages}{3--29}.

\bibitem[{\citenamefont{Mandelbrot}(1990)}]{Mandelbrot-1990-PA}
\bibinfo{author}{\bibfnamefont{B.~B.} \bibnamefont{Mandelbrot}},
  \bibinfo{journal}{Physica A} \textbf{\bibinfo{volume}{163}},
  \bibinfo{pages}{306} (\bibinfo{year}{1990}).

\bibitem[{\citenamefont{Mandelbrot}(1991)}]{Mandelbrot-1991-PRSLA}
\bibinfo{author}{\bibfnamefont{B.~B.} \bibnamefont{Mandelbrot}},
  \bibinfo{journal}{Proc. Roy. Soc. London A} \textbf{\bibinfo{volume}{434}},
  \bibinfo{pages}{79} (\bibinfo{year}{1991}).

\bibitem[{\citenamefont{Chhabra and
  Sreenivasan}(1991)}]{Chhabra-Sreenivasan-1991-PRA}
\bibinfo{author}{\bibfnamefont{A.~B.} \bibnamefont{Chhabra}} \bibnamefont{and}
  \bibinfo{author}{\bibfnamefont{K.~R.} \bibnamefont{Sreenivasan}},
  \bibinfo{journal}{PRA} \textbf{\bibinfo{volume}{43}}, \bibinfo{pages}{1114}
  (\bibinfo{year}{1991}).

\bibitem[{\citenamefont{Zhou et~al.}(2001)\citenamefont{Zhou, Liu, and
  Yu}}]{Zhou-Liu-Yu-2001-Fractals}
\bibinfo{author}{\bibfnamefont{W.-X.} \bibnamefont{Zhou}},
  \bibinfo{author}{\bibfnamefont{H.-F.} \bibnamefont{Liu}}, \bibnamefont{and}
  \bibinfo{author}{\bibfnamefont{Z.-H.} \bibnamefont{Yu}},
  \bibinfo{journal}{Fractals} \textbf{\bibinfo{volume}{9}},
  \bibinfo{pages}{317} (\bibinfo{year}{2001}).

\bibitem[{\citenamefont{Zhou and Yu}(2001{\natexlab{a}})}]{Zhou-Yu-2001-PA}
\bibinfo{author}{\bibfnamefont{W.-X.} \bibnamefont{Zhou}} \bibnamefont{and}
  \bibinfo{author}{\bibfnamefont{Z.-H.} \bibnamefont{Yu}},
  \bibinfo{journal}{Physica A} \textbf{\bibinfo{volume}{294}},
  \bibinfo{pages}{273} (\bibinfo{year}{2001}{\natexlab{a}}).

\bibitem[{\citenamefont{Zhou and Yu}(2001{\natexlab{b}})}]{Zhou-Yu-2001-PRE}
\bibinfo{author}{\bibfnamefont{W.-X.} \bibnamefont{Zhou}} \bibnamefont{and}
  \bibinfo{author}{\bibfnamefont{Z.-H.} \bibnamefont{Yu}},
  \bibinfo{journal}{Phys. Rev. E} \textbf{\bibinfo{volume}{63}},
  \bibinfo{pages}{016302} (\bibinfo{year}{2001}{\natexlab{b}}).

\bibitem[{\citenamefont{Olsen}(1998)}]{Olsen-1998-PMH}
\bibinfo{author}{\bibfnamefont{L.}~\bibnamefont{Olsen}},
  \bibinfo{journal}{Periodica Methematica Hungaria}
  \textbf{\bibinfo{volume}{37}}, \bibinfo{pages}{81} (\bibinfo{year}{1998}).

\bibitem[{\citenamefont{Olsen}(1999)}]{Olsen-1999-HMJ}
\bibinfo{author}{\bibfnamefont{L.}~\bibnamefont{Olsen}},
  \bibinfo{journal}{Hiroshima Math. J.} \textbf{\bibinfo{volume}{29}},
  \bibinfo{pages}{435} (\bibinfo{year}{1999}).

\bibitem[{\citenamefont{Olsen}(2000)}]{Olsen-2000-PP}
\bibinfo{author}{\bibfnamefont{L.}~\bibnamefont{Olsen}},
  \bibinfo{journal}{Progress in Probability} \textbf{\bibinfo{volume}{46}},
  \bibinfo{pages}{3} (\bibinfo{year}{2000}).

\end{thebibliography}

\end{document}